# Deconstructing Bataknese *Gorga* Computationally

Hokky Situngkir
[hokky.situngkir@surya.ac.id]
Dept. Computational Sociology
Bandung Fe Institute

**Abstract**

The carved and painted decorations in traditional Batak houses and buildings, *gorga*, are the source of their exoticism. There are no identical patterns of the ornaments within Batak houses and the drawings are closely related to the way ancient Batak capture the dynamicity of the growing "tree of life", one of central things within their cosmology and mythology. The survey of ornaments of Batak houses and buildings in Northern Sumatera Indonesia has made us possible to observe the complex pattern. The fractal dimensions of the geometrical shapes in *gorga* are calculated and they are conjectured into 1.5-1.6, between the dimensional of a line and a plane. The way *gorga* is drawn is captured by using some modification to the turtle geometry of L-System model, a popular model to model the dynamics of growing plants. The result is a proposal to see Bataknese *gorga* as one of traditional heritage that may enrich the studies to the generative art.

# 1. Introduction

The Indonesian Bataknese decorates their traditional houses and buildings with carved or painted ornaments known as "*gorga*". The touch of the traditional knowledge with modernity has even yielded many modern buildings in Northern Sumatera, Indonesia, decorated with the traditional *gorga*, for example hotels, churches, governmental offices, community buildings, *et cetera,* expressing the identity of Batak people throughout the country. However, while the traditions have been slowly declined, we are still lack of studies due to the geometry of the ornamentation within *gorga*.

Etymologically speaking, the word "*gorga*" refers to the bataknese ornaments, motifs, of decoration; most of them are carved (*uhir*) or painted (*dais*) on woods, the elements of traditional buildings. The variations of Bataknese *gorga* are so many, be it geometrically, shapes and stylistic implemented into the woods. Some observers have made classifications of *gorga* based on the geometric shape emanates from the carved woods. Some *gorga* is in shape of human being, and it is called *gorga tarus*. Some *gorga* express the shape of animals, like horses (*hoda-hoda*), lizard (*boraspati* symbols of land fertility), ox (*sijonggi*), and some others show the shape of giants or monsters, for instance the shape of lion (*singa-singa*), head of buffalo (*ulu paung*), etc. The shapes of flora and other natural objects are also shown in *gorga*, for example the tree symbolizing life (*hariara sundung di langit*), some kind of moss (*simeol-eol*), the sun (*simataniari*), turbulent water flow (*silintong*), teeth of animals (*ipon-ipon*), the traditional batak compass (*desa naualu*), *et cetera*. All of the *gorga* decorate in three colors, black, red, and white. The batak community believes that any shapes of the *gorga* represent some philosophical meanings related to the way of life of batak traditional people [4].

The motifs of bataknese *gorga* are somehow related to the mythologies of traditional communities and the modern living might have made the relations have long forgotten. Most of batak drawings represent the traditional cosmology of the tree of life [5]. While great parts of bataknese *gorga* are in the shape of curlicues, the meanings are mentioned as portrayal of the dynamics of fluid, the growing plants, and many more due to the cosmology of the a tree, named the tree of life [17]. The geometric style of drawings of the trees, with its branches, roots, leaves, *et cetera*, are somehow also in some drawings of *pustaha*, the traditional books and manuscripts of batak people [18]. It is possible that the drawings we can find in the bataknese houses are in the aesthetic form of the cosmology (*hadatuon*) of ancient batak as we could browse through some dedicated catalogue of the ancient manuscript [19]. Into the ancient batak sclupures, the similar patterns are also implemented as well [10]. Those sculptures are closely related to ancient spiritualism and religions of ancient Batak community.

In fact, such 'drawings' are also in parallel discovered in the traditional ornamentations and decorations from other places in Indonesia [20]. The varieties of the way traditional people depicting and symbolizing the growing plants, the dynamics of fluids, clouds, and so on have interesting similarities [15]. Hypothetically speaking, our endeavor to reveal the geometric shape of bataknese *gorga* might deliver us into some new way seeing the diverse similar models throughout the Indonesian archipelago.

The paper reports some scrutinizing effort on the "geometry" of most elements of *gorga*, which is related to the portrayal of the growing elements, be it branches, leaves, or anything within the tree of life, yet somehow have been named as other objects, like moss, flow of water, *et cetera* accordingly. The data are from houses and buildings with *gorga* ornaments scattered in near the Lake Toba, North Sumatera, Indonesia. Some examples of the collected data are shown in figure 1.

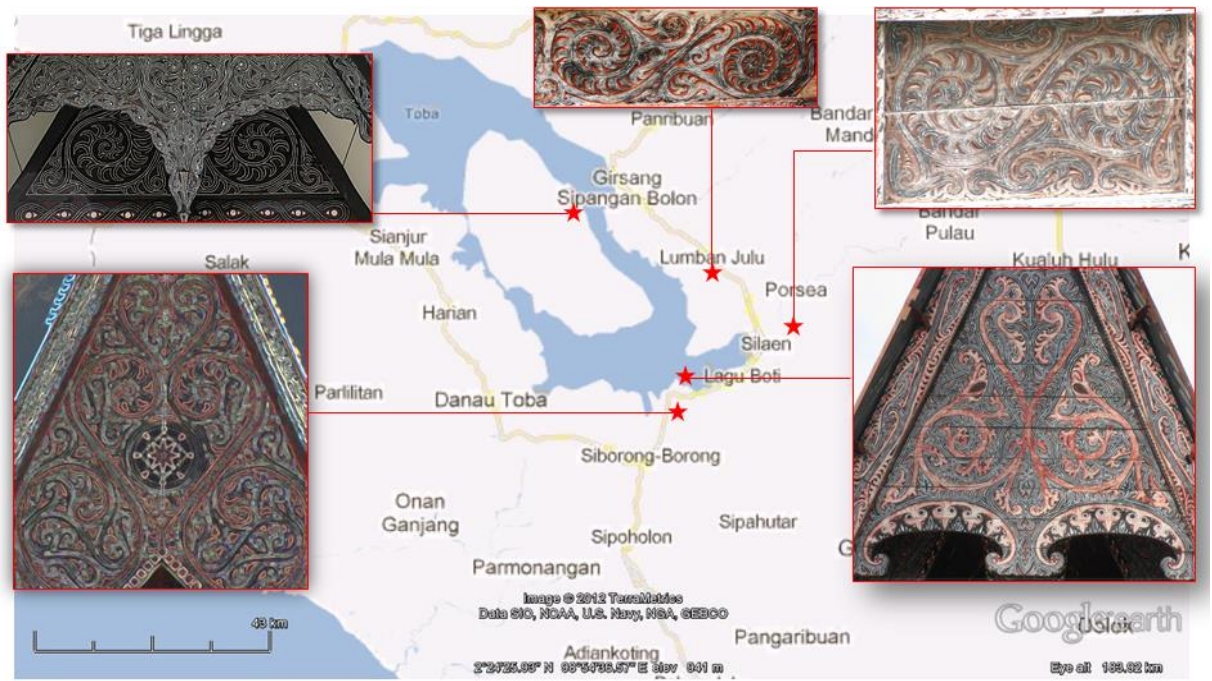

**Figure 1.** Some variations of *gorga* carved in woods as decorations in traditional houses from some places in North Sumatera, Indonesia, clockwise from the left-top: House in Raja Sidabutar tomb, Jangga Dolok traditional village, Janji Maria traditional village, Building of Traditional marketplace in Balige, and Building of Municipal Office in Toba Samosir.

The paper is structured as follow. In the next section, we discuss the result of calculation of the dimensionality of *gorga* batak collected from houses and buildings. The result shows the fractal dimension of hundreds of bataknese *gorga* we collected. This findings are thus become our starting point of capturing the rules incorporated on making the *gorga*. Here, *gorga* becomes a kind of generative art which can utilize the computational processes. The following section discuss the turtle geometry, a modern computational tool for capturing the geometry of the growing plants referred to the so called L-System.

## 2. The dimensionality of *gorga*

When *gorga* is expected to capture the growing dynamics of natural objects like trees, branches, water flow, and so on, then the expectation is to have a portrait of fractals [*cf.* 2]. Discussions in [14] have shown us how ancient Javanese did similar thing with their batik. It is interesting to see how a lot of ornamentations within Indonesian complex traditional crafts [11] are made upon the acquisition of simple rules generatively, which in turn deliver us an interesting objects to be observed in the sense of fractal geometry [12].

We know that fractal dimension measures the self-similarity of the object or any quantities: the similar structures at different scales at the object. The measurement we use here is the box-counting (Mikonwski) dimension, *i.e.:* the fractal dimension of a set of $S$ in metric or Euclidean space, $R^N$. If we place the object into the observational space with evenly-spaced grids and count how many boxes are required to cover the object or set, the fractal dimension is then calculated by counting the changes of the boxes count as finer grid is applied. This is illustrated in figure 2, a Bataknese *gorga* and the respective result of the box counting. The set is fractal as the calculation gives a power-law,

$$N \sim r^{-D} \tag{1}$$

Here, we have *D* as the fractal (capacity) dimension, or more formally [16], that

$$D = \lim_{r \to 0} \frac{\log[\frac{1}{N}]}{\log r} \tag{2}$$

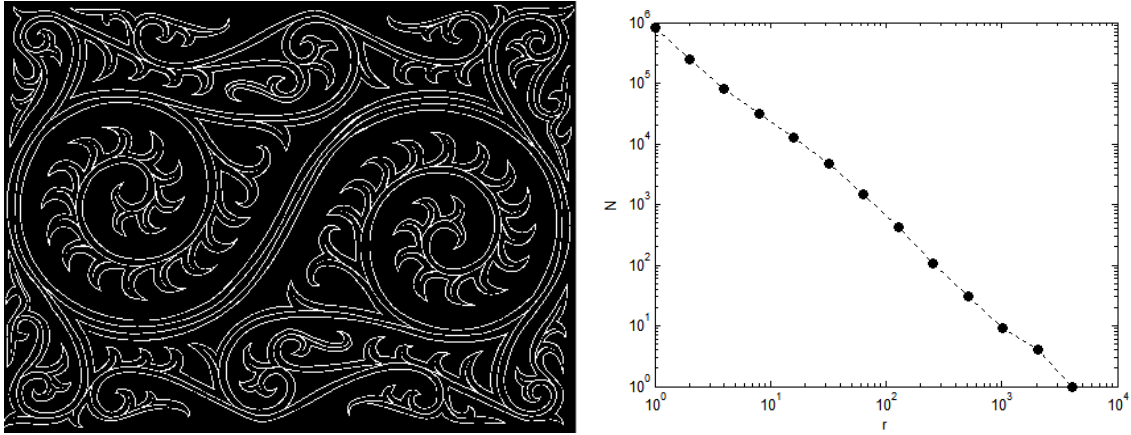

**Figure 1.** A *gorga* and the respective box counting calculation.

Few results of our box counting calculation to the *gorga* are shown in table 1. It is interesting to find that the dimensions are at the average of 1.5-1.6. This kind of geometric shapes remind us to some famous fractals with the similar dimension, for example the dragon curve ($D = 1.52$) and Sierpinski triangle ($D = 1.58$) [7]. The dimensions are between the dimension of a line, which dimension is 1, and a 2 dimensional plane; *gorga* is not a line, nor a plane, but in between.

*Gorga* could be seen as shape of curlicue filling up the space of with paintings or carvings. The patterns are complex yet the makers of *gorga* should follow simple steps for the drawings. It is obvious that the logic used by the makers of *gorga* has similarities with those the makers of *batik* in Java Island. The traditional living have not yet be introduced with the metric system to make such complex shapes and patterns, but the rule is simple. The simplicity yields the complexity, a seemingly random curlicues but not random actually. The constraint is simple, how to make drawings to fill the "emptiness" of the space, a rather similar process that we can capture with the visual geometry of turtle [1].

**3. The turtle geometry of *gorga***

When it comes to model the growing multi-cellular plants, Lindenmayer proposed L-System in 1968. The L-System is a proposal to see how things grow from the initial conditions, and by following the stated rules, the complex patterns and shapes are emerged from the rewriting process [8].

An L-System can be denoted geometrically as $G = \langle V, \omega, P \rangle$. $V$ is the alphabet within the system with $V^*$ as the words, and $V^+$ is the set of the non-empty one over the system. Within the system is also the non-empty words axiom $\omega \in V^+$, and the finite set of productions $P \subset V \times V^*$. For instance, the production $(a, \chi) \in P$ is written for the production of word $\chi$ from the letter $a$ in the previous iteration. The assumption of the L-System is that for any $a \in V$, there at least one word $\chi \in V^*$ such that $a \to \chi$.

**Table 1**
Some *gorga* and the estimated box-counting dimensions

| Gorga | Estimated Box-counting Dimension |
|---|---|
| 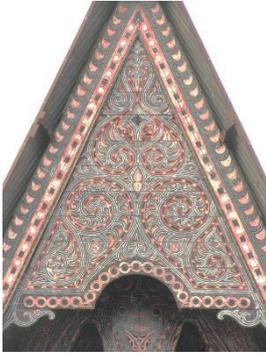 | 1.5348 ± 0.23374 |
| 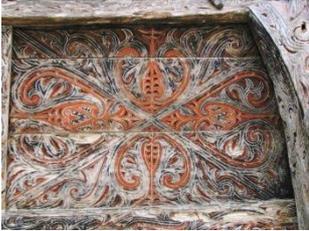 | 1.5165 ± 0.21393 |
| 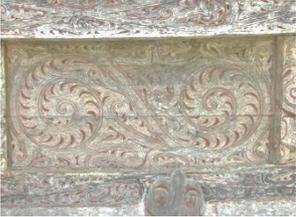 | 1.4864 ± 0.21325 |
| 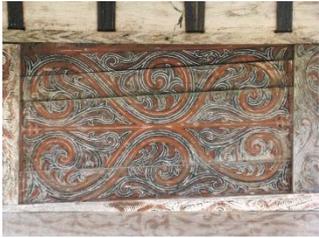 | 1.5614 ± 0.23688 |
| 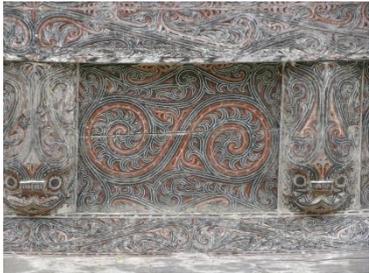 | 1.5924 ± 0.24983 |

In the computational implementation, the L-System is represented by the turtle geometry. A turtle is defined by its current position and heading $(x, y, \alpha)$ with incremental step size $d$ and angle $\delta$. Thus we can write the local movement of the turtle with alphabet,

$F : (x, y, \alpha) \rightarrow (x', y', \alpha)$, where $x' = x + d\cos\alpha$ and $y' = y + d\sin\alpha$ while drawing down the movement.

$+ : (x, y, \alpha) \rightarrow (x, y, \alpha + \delta)$

$- : (x, y, \alpha) \rightarrow (x, y, \alpha - \delta)$

$[$ : push the current state of the turtle, e.g.: $(x, y, \alpha)$ into the stack.

$]$ : pop the state of the turtle, e.g.: $(x, y, \alpha)$ from the stack into the current state.

The L-System and its derivation for many visual geometry is established from the above basic definitions, be it the visual model of the plants and multi-cellular objects and fractals [9], including the way we would observe the Bataknese *gorga*. This approach is somehow similar with a previous work shown in [3] on explaining the *origami* and *kirigami*.

The basic variations of *gorga* simply could be emerged by implementing the turtle geometry with at least three production rules from simple axiom,

$\omega : F$

$p_{random} : F \rightarrow \|F\|\|F\|\|F\|[-random(\delta)\|F\|] + \|F\| + \|F\|$

$p_{angle} : F \rightarrow \|F\|\|F\|\|F\|[-(\delta)\|F\|] + \|F\| + \|F\|$

$p_{spiral} : F \rightarrow \|F\|\|F\|\|F\|[-\tau(\delta)\|F\|] + \|F\| + \|F\|$

We introduce a new notation $\|...\|$, which means that the incremental step within would only be delivered while there is no potentially overlapping with any previously drawings. This is the major difference of *gorga* and standard L-System: the production rule is space filling with a local feedback that no overlapping is allowed. This explains the variations of *gorga* across Bataknese traditional houses and buildings decorated with *gorga* in northern Sumatera.

Some *gorga* can be produced by seemingly random production rule ($p_{random}$). We call it "seemingly random" for the randomness is led by the local observation of the turtles while moving along with the rule: it is heading to the least local places have been drawn previously.

Some other *gorga* are produced by having the turtles move with some constant angle ($p_{angle}$) or with gradual changes of angle ($\delta$) yielding the simple spiraling *gorga* ($p_{spiral}$). Both curlicues are drawn by the turtles with local constraint not overlap any drawing from the previous rounds. After running rounds of simulations, the modified L-System would eventually emerge a pattern that might have similar geometry with Bataknese *gorga*.

In order to utilize the L-System to capture the dynamics of plants' growth, some modification to the original L-System is modified into the parametric L-System [9], and in return, in order to capture the way *gorga* makers' way of thinking while carving and painting *gorga*, the modification to the parametric L-System is delivered. This process is illustrated in figure 3. The symbol "◊" in the flowchart represents the process of observing the localities in order not to overlap the previous drawing.

The geometry described in this section is implemented in the logo-like turtle, and we could positively see how *gorga* has been a kind of generative art from the traditional Bataknese wisdom.

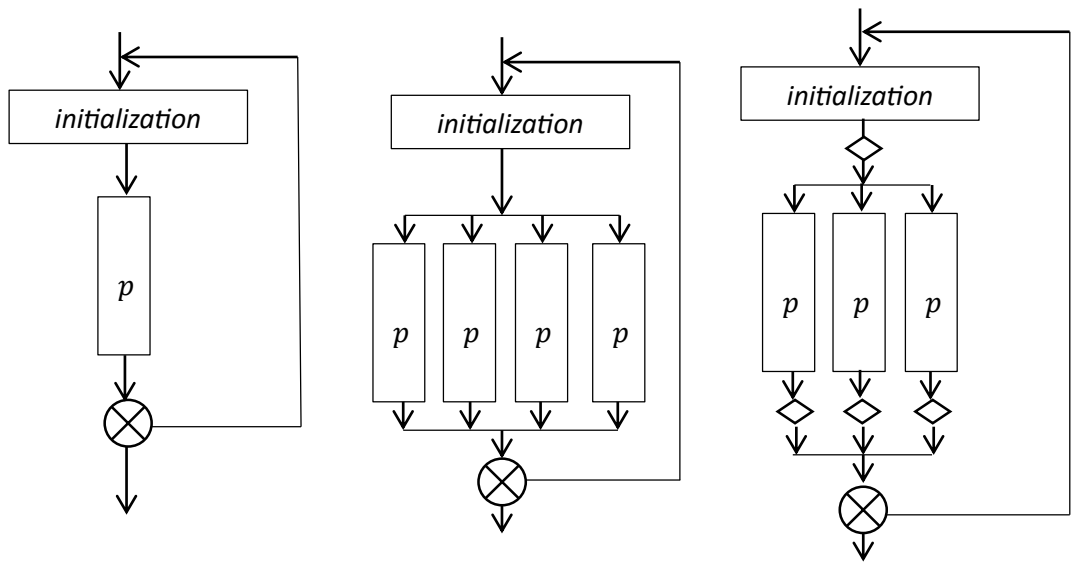

**Figure 3.** The difference of turtle geometry among classical L-System (*left*), parametric L-System (*middle*), and the implementation of *gorga* (*right*).

Following the simple rules yet emerging the complex pattern is an interesting feature of Indonesian traditional arts and crafts. The simplicity is a thing for the lack of standard modern abstractions about space, standardization on measurement; in short the elements of modern geometry. It is interesting, nonetheless, that when modern computational geometry focus on the emergence of complexity from simple processes, we can capture the way the ancient civilization exhibiting complex patterns in their cultural artifacts [13]. Eventually, we can see how the innovative scientific endeavor sees the Bataknese *gorga,* for instance, was perhaps a kind of generative art. When generative art is a new kind of doing arts, a meeting point is there with one we have since long time ago when computation is nothing but organic processes [6].

This is what we can do with the previous sections. We do some customizations for productions of art based on the turtle geometry of Bataknese *gorga* in order to have a computational tool to simulate our findings about the basic characteristics of *gorga*,

- The algorithmic process is to fill the available space; practical traditionally can be made upon carving and paintings on woods.
- Exploiting the simple patterns of the growing plantations, i.e.: curlicues, in order to have the aesthetic aspects of decorations.

The result is a computational program that can emerge the two aspects by employing the turtle geometry of *gorga*.

One important aspect is that the computational drawing should be able to bring the algorithmic generative drawings in some particular shapes, be it ellipse, circles, rectangles, triangles, and so on. The rules of production are then implemented upon the basic shapes of the space to be filled with the generative drawings. Some results of computational simulation incorporating the production rules are shown in figure 4.

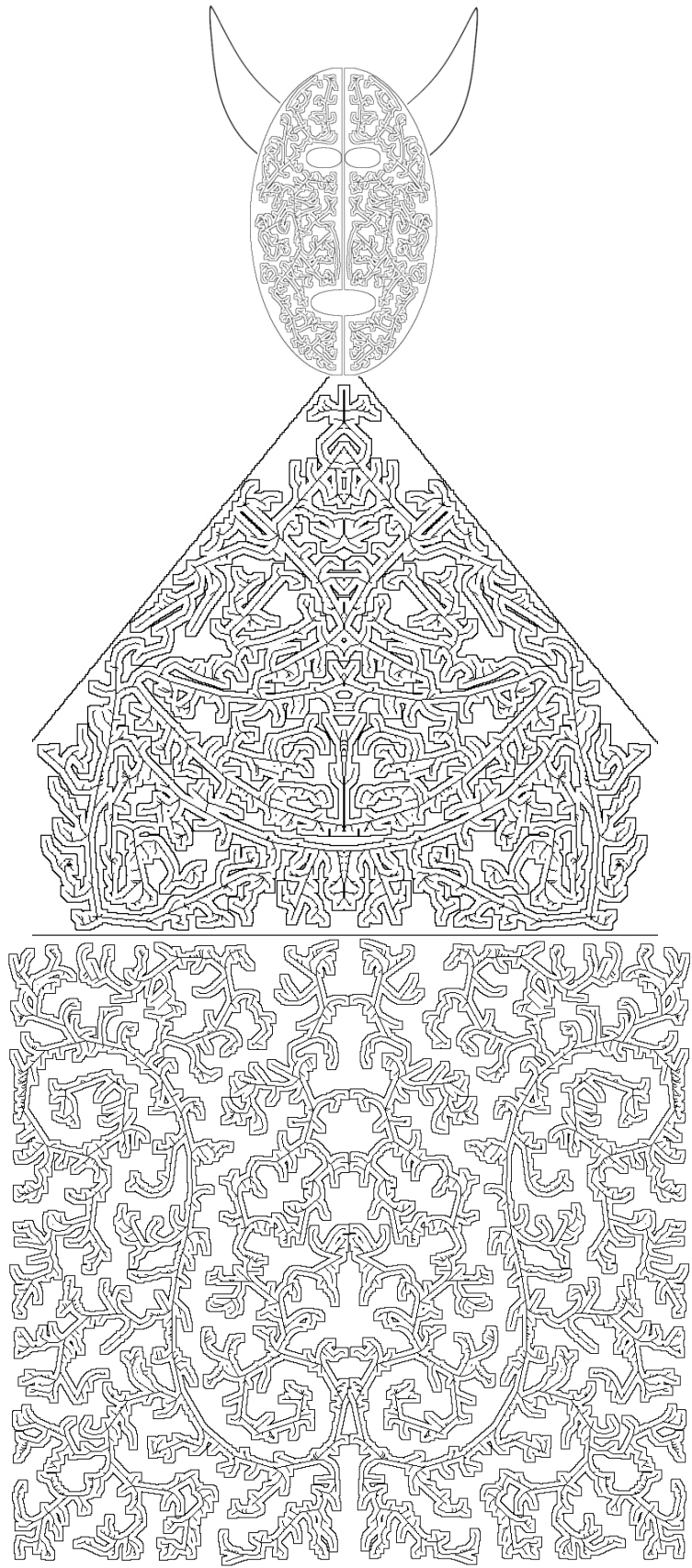

**Figure 4.** Examples of the result of computational simulations upon three different shape of drawing spaces: rectangle (*bottom*), triangle (*middle*), and ellipse (*top*).

## 4. Concluding Remarks

The Bataknese *gorga* carved and/or painted on woods are unique decorative ornaments widely found in traditional houses and buildings showing the social identity of heritage from generations to generations among Batak community in Indonesia. The complexity of the ornaments are puzzling for they are rooted to some ancient philosophy, cosmology, and wisdom from the past, yet aesthetically intriguing for its precisions filling out the decorative space on the elements of buildings, houses, and even traditional sculptures. From the documentation from buildings and traditional houses around the Lake Toba, North Sumatera, Indonesia, we do a survey showing the fractal geometry emanated from the traditional ornaments. The ornaments expect to capture the dynamics of the growing patterns of plantations, flowing fluid, and so on, which is naturally fractals.

By crunching the ways the traditional people make *gorga*, we propose a modification of turtle geometry that may be utilized to mimic how the simple rules emerge the complex patterns of *gorga*. There are no identical *gorga* across the houses and buildings in Northern Sumatera, yet the simple rules that the makers follow may identical. The basic principles are the constraint to fill the space and at the same time trying to capture the dynamicity of growth, a concept that is important in the ancient Batak cosmology and philosophy.

The turtle geometry and the characteristics of fractals within *gorga* are thus implemented computationally and interestingly demonstrate how Bataknese *gorga* can be seen as a kind of generative art. The emerging of seemingly random aesthetic order from the simple rules of production is the main course of the computational process.

The computation has helped us to appreciate more on the element of traditional culture, in this case, Bataknese *Gorga.*

**Acknowledgement**

I thank Vande Leonardo (Indonesian Archipelago Cultural Initiatives) for the expedition of documentation of *gorga* in Northern Sumatera, Bungaran Simanjuntak for friendly sessions of discussion about Bataknese *gorga,* Usman Apriadi for works on rastering the *gorga* images, and Surya Research International for the support in which period the research is delivered.